# The Informational Architecture Of The Cell

Sara Imari Walker[1,2,3], Hyunju Kim[2] and Paul C.W. Davies[2]

[1] *School of Earth and Space Exploration, Arizona State University, Tempe, AZ 86281, USA*
[2] *Beyond Center for Fundamental Concepts in Science, Arizona State University, Tempe, AZ 86281, USA*
[3] *Blue Marble Space Institute of Science, Seattle WA USA*

# Abstract

We compare the informational architecture of biological and random networks to identify informational features that may distinguish biological networks from random. The study presented here focuses on the Boolean network model for regulation of the cell cycle of the fission yeast *Schizosaccharomyces Pombe*. We compare calculated values of local and global information measures for the fission yeast cell cycle to the same measures as applied to two different classes of random networks: Erdos-Renyi and Scale-Free. We report patterns in local information processing and storage that do indeed distinguish biological from random, associated with control nodes that regulate the function of the fission yeast cell cycle network. Conversely, we find that integrated information, which serves as a global measure of "emergent" information processing, does not differ from random for the case presented. We discuss implications for our understanding of the informational architecture of the fission yeast cell cycle network in particular, and more generally for illuminating any distinctive physics that may be operative in life.

# 1. Introduction

Although living systems may be decomposed into individual components that each obey the laws of physics, at present we have no explanatory framework for going the other way around - *we cannot derive life from known physics*. This, however, does not preclude constraining what the properties of life must be in order to be compatible with the known laws of physics. Schrödinger was one of the first to take such an approach in his famous book *"What is Life?"* [1]. His account was written prior to the elucidation of the structure of DNA. However, by considering the general physical constraints on the mechanisms underlying heredity, he correctly reasoned that the genetic material must necessarily be an "aperiodic crystal". His logic was two-fold. Heredity requires stability of physical structure – hence the genetic material must be rigid, and more specifically crystalline, since those are the most stable solid structures known. However, normal crystals display only simple repeating patterns and thus contain little information. Schrödinger therefore reasoned that the genetic material must be aperiodic in order to encode sufficient information to specify something as complex as a living cell in a (relatively) small number of atoms. In this manner, by simple and general physical arguments, he was able to accurately predict that the genetic material should be a stable molecule with a non-repeating pattern – in more modern parlance, that it should be a molecule whose information content is *algorithmically incompressible*.

Today we know Schrödinger's "aperiodic crystal" as DNA. Watson and Crick discovered the double helix just under a decade after the publication of *"What is Life?"* [2]. Despite the fact that the identity of the genetic material was discovered over sixty years ago, we are in many ways no closer today than we were before its discovery to having an explanatory framework for biology. While we have made significant advances in understanding biology over the last several decades, *we have not made comparable advances in physics to unite our new understanding of biology with the foundations of physics.* Schrödinger used physical principles to constrain unknown properties of biology associated with genetic heredity. One might argue that this could be done again, but now at the level of the epigenome, or interactome, and so on for any level of biological organization, and not just for the genome as Schrödinger did. But this kind of approach only serves to constrain structures consistent with the laws of physics; it does not explain why they should exist in the first place.[1] An alternative approach is to instead use insights from biology to constrain unknown physics. That is, to take a track working in the opposite direction from that proposed by Schrödinger: rather than using physics to inform biology, one could start by thinking about *biology as a means to inform potentially new physics,* if indeed such missing physics exists. This is the approach we take here.

*Author for correspondence: Sara Imari Walker (sara.i.walker@asu.edu).
†Present address: School of Earth and Space Exploration,
Arizona State University, Tempe, AZ 86281, USA

The challenge with such an approach is that it is not immediately evident what physics might explain life: if it were an easy connection, such that life immediately dropped out of the known laws of physics, the problem would surely have been solved in the 70 years since Schrödinger's seminal book, or earlier. A problem is that we do not know what features might distinguish biological networks from other classes of complex physical systems. Many advances in physics have been preceded by elucidating the physical structure of a simple example of the class of systems under investigation: for example, the Carnot engine provided deep insights into the nature of thermodynamics and likewise, understanding the structure of the hydrogen atom provided important insights into atomic and quantum physics. What would be an analogous structure to study for biology? And what should one do to attempt to characterize it? There is widespread regard that information, in as yet unspecified capacity, may hold the clue for finally answering Schrödinger's question [3]. We therefore direct our attention to information-processing systems in biology, and specifically to biological networks, which resemble in many ways electronic or computational information-processing systems [4]. Since biology is dauntingly complex, we follow a long tradition in physics to gain deeper insights into physical structure: identify the simplest system(s) that can be described, and utilize idealized minimal mathematical representations that still capture the essential qualities under investigation – in this case, life's informational structure.

One of the simplest biological networks known that still accurately captures biological function is the cell cycle regulatory network of the fission yeast *Schizosaccharomyces Pombe* [5], which we regard as a sort-of "hydrogen atom" for complex information-rich biological networks in the current study. The essential aspects of this system for our analysis are well represented as a Boolean network in which key regulator proteins for the cell-cycle process are represented as either "on" (1) or "off" (0), and can activate or inhibit other proteins in the network. Our approach is to compare the informational architecture of this Boolean network model to ensembles of random networks to identify informational properties that might distinguish the biological network from random. For this purpose, we utilize two different classes of random networks: Erdos-Renyi (ER) random networks and Scale-Free (SF) random networks. The ER networks share the same total number of nodes and globally the same total number of activation and inhibition links as the biological network, but are otherwise fully randomized over all network topologies. The SF networks are more constrained, maintaining the same number of activation and inhibition links for each individual node. The SF networks therefore share important topological properties with the biological fission yeast cell cycle network, including degree distribution rank ordering the number of edges per node.

---

[1] There is an interesting point here in the kind of prediction made by Schrödinger – he was able to accurately predict the general structure of the genetic material, but only by first taking as an axiom that hereditary structures exist that can reliably propagate sufficient "information" to specify an organism from one generation to the next. So his "first principles" argument already makes strong assumptions about physical reality and the existence of "life-like" physical systems.

It is unclear at present precisely which concept(s) of "information" are most relevant to biological organization (see *e.g.* [6] for a review of the philosophical debate on the ontological status of the concept of information in biology). We therefore consider several local and global measures in what follows, adopted from information dynamics [7-9] and integrated information theory [10,11], respectively. Features shared by the biological network and SF networks, but not the ER networks can be concluded to arise as a result of topological features, whereas those observed in the biological network that are not shared with the SF networks should be regarded as arising specifically due to network features distinctive to biological function. By comparing results for the biological network to both ER and SF random networks, we are therefore able to distinguish which features of the informational architecture of biological networks arise as a result of network topology (*e.g.,* degree distribution, which is shared with the SF networks but not the ER networks) and which are peculiar to biology (and presumably generated by the mechanism of natural selection). By implementing the local measures of information processing and information storage provided by information dynamics, we find patterns in local information processing and storage that do indeed distinguish the biological from either the SF or ER random networks, associated with regulation of the function of the fission yeast cell cycle network by a subset of control nodes. Conversely however, we find that integrated information, which serves as a global measure of "emergent" information processing, does *not* differ from random for the case presented. We discuss implications for our understanding of the physical structure of the fission yeast cell cycle network based on the uncovered informational architecture, and a look forward toward illuminating any distinctive physics operative in life.

## 2. A Phenomenological Procedure for Mapping the Informational Landscape of Biological Networks

In this work, we address the foundational question of how non-random patterns of informational architecture uncovered in biological networks might offer general insights into the nature of biological organization. Henceforth we use the term *"informational architecture"* rather than "information" or "informational pattern", because architecture implies the constraint of a physical structure whereas patterns in the abstract are not necessarily tied to their physical instantiation (see also [12]). We focus our analysis on a Boolean network model for a real biological system – specifically the cell cycle regulatory network of the fission yeast *Schizosaccharomyces Pombe (S. Pombe)* – which has been demonstrated to accurately model cell cycle function [5]. A motivation for choosing this network for our study is that it is small, and accurately models what is arguably one of the most essential biological functions: regulation of cellular division. With just nine nodes the network is computationally tractable for all of the information theoretic measures we implement in our study – including the computationally-intensive integrated information [11]. The fission yeast cell cycle network also shares important features with other Boolean network models for biological systems, including the shape of its global attractor landscape and resultant robustness properties [5], and the presence of a control kernel (described below) [13]. It therefore serves our purposes as a sort-of "hydrogen atom" for complex information-rich systems. Although we focus here on this simple gene regulatory network, we note that *our analysis is not level specific*. The formalism introduced is intended to be universal and may apply to networks any level of biological organization from the first self-organized living chemistries all the way up to cities and technological societies.

We note that although this special theme issue is focused on the concept of "DNA as information", we do not explicitly focus on DNA *per se* herein. We instead consider distributed information processing as it operates within the cell, as we believe such analyses have great potential to explain, for example, why physical structures capable of storage and heredity, such as DNA, should exist in the first place. Thus, we regard an explanation of "DNA as information" to arise only through a proper physical theory that encompasses living processes, which should be illuminated by the kinds of analyses presented here.

## 2-1) Fission Yeast Cell-Cycle Regulatory Network: A Case Study in Informational Architecture

Regulation of cellular division is a central aspect of cellular function. In the fission yeast *S. Pombe* the cell passes through a number of phases, G1 – S – G2 – M, which collectively constitute its cell-cycle dictating the steps of cellular division to produce two daughter cells from a single parent cell. During the G1 stage, the cell grows, and if conditions are favourable, it can commit to division. During the S stage, DNA is replicated. In the G2 stage, there is a "gap" between DNA replication (in the S phase) and mitosis (in the M phase) where the cell continues to grow. In the M stage, the cell undergoes mitosis and two daughter cells are produced. After the M stage, the daughter cells enter G1 again, thereby completing the cycle.

The biochemical reactions that form the network controlling the cell cycle for *S. Pombe* have been studied in detail, and a Boolean network model has been constructed that has been shown to accurately track the phases of cellular division for *S. Pombe* (see [5] and references therein). The interaction graph for the Boolean network model is shown in Fig. 1. Each node corresponds to a protein needed to regulate cell cycle function. Nodes may take on a Boolean value of '1' or '0', indicative of whether the given protein is present or not at a particular step in the cycle (labels indicate the relevant proteins). Edges represent *causal* biomolecular interactions between proteins, which can either activate or inhibit the activity of other proteins in the network (or themselves). In the model, the successive states $S_i$ of node $i$ are determined in discrete time steps by the updating rule:

(1)
$$S_i(t+1) = \begin{cases} 1, & \sum_j a_{ij} S_j(t) > \theta_i \\ 0, & \sum_j a_{ij} S_j(t) < \theta_i \\ S_i(t), & \sum_j a_{ij} S_j(t) = \theta_i \end{cases}$$

where $a_{ij}$ denotes weight for an edge $(i, j)$ and $\theta_i$ is threshold for a node $i$. The threshold for all nodes in the network is 0, with the exception of the proteins Cdc2/13 and Cdc2/13*, which have thresholds of −0.5 and 0.5, respectively in the Boolean network model for *S. Pombe*. For each edge, a weight is assigned according to the type of the interaction: $a_{ij}(t) = -1$ for inhibition and $a_{ij}(t) = +1$ for activation, and $a_{ij}(t) = 0$ for no direct causal interaction. This simple rule set captures the causal interactions necessary for the regulatory proteins in the fission yeast *S. Pombe* to execute the cell cycle process. Although many of the fine details of biological complexity (such as kinetic rates, signalling pathways, noise, asynchronous updating) are jettisoned by resorting to such a coarse-grained model, it does retain the key feature of *causal architecture* necessary for our analysis presented here.

## 2-2) The Dynamics of the Fission Yeast Cell-Cycle Regulatory Network

In the current study, we will make reference to informational attributes of both individual *nodes* within the network, and the *state* of the network as a whole, which is defined as the collection of all Boolean node values (*e.g.*, at one instant of time). We study both since we remain open-minded about whether informational patterns potentially characteristic of biological organization are attributes of nodes or of states (or both). When referring to the *state-space* of the cell cycle we mean the space of the $2^9 = 512$ possible states for the nine-node network as a whole. We refer to the *global* causal architecture of the network as a mapping between network states (Fig. 2), and the *local* causal architecture as the edges (activation or inhibition) within the network (Fig. 1).

*Time Evolution of the Fission Yeast Cell-Cycle Network.* Iterating the set of rules in Eq. 1 accurately reproduces the time sequence of network states corresponding to the phases of the fission yeast cell cycle, as measured *in vivo* by the activity level of proteins (see [5] for details). Initializing the network in each of the 512 possible states and evolving each according to Eq. (1) yields a flow diagram of network states that details *all* possible dynamical trajectories for the network, as shown in Fig. 2. The flow diagram highlights the global dynamics of the network, including its attractors. In the diagram, each point represents a unique pattern of Boolean values for individual nodes corresponding to a network state. Two network states are connected if one is a cause or effect of the other. More explicitly, two network states $G$ and $G'$ are causally connected when either $G \to G'$ ($G$ is maps to, or is a *cause* for $G'$) or $G' \to G$ ($G'$ instead is the cause, and $G$ the effect) when the update rule in Eq. 1 is applied locally to individual nodes. The notion of *network states* being a cause or an effect may be accommodated within integrated information theory [10; 11], which we implement below for the fission yeast cell cycle network. We note that because the flow diagram contains all mappings between network states it captures the entirety of the *global causal structure* of the network, encompassing any possible state transformation consistent with the local rules in Eq. 1.

For the fission yeast cell cycle network, each initial state flows into one of sixteen possible attractors (fifteen stationary states and one limit cycle). The network state space can be sub-divided according to which attractor each network state converges to, represented by the colored regions in the left-hand panel of Fig. 2. About 70% of states terminate in the primary attractor shown in red [5]. This attractor contains the biological sequence of network states corresponding to the four stages of cellular division: G1—S—G2—M, which then terminates in the inactive G1 state. The model therefore directly maps the *function* of the cell cycle to the dynamics of its Boolean network representation.

*The Control Kernel Nodes: Regulators of Network Function.* An interesting feature to emerge from previous studies of the fission yeast cell cycle network, and other biologically motivated Boolean network models, is the presence of a small subset of nodes – called the control kernel – that governs global dynamics within the attractor landscape [13]. When the values of the control kernel are fixed to the values corresponding to the primary attractor associated with biological function, the entire network converges to that attractor regardless of the initial state of the network (see Fig. 2, right panel) – that is, *the control kernel nodes regulate the function of the network*. Here, function is defined in terms of dynamics on the attractor landscape as noted above. The control kernel of the fission yeast network is highlighted in red in Fig. 1. As we will show, it plays a key role in the informational architecture of the fission yeast cell cycle.

## 2-3) Constructing Random Networks for Comparison to Biological Networks

We compare the informational architecture of the fission yeast cell cycle Boolean network to two classes of random networks: Erdos-Renyi (ER) and Scale-Free (SF)[2]. Both classes retain certain features of the biological network's causal structure as summarized in Table 1, and described in detail by us in [12] (see also [14; 15; 16]). Both ensembles of random networks share the same number of nodes and the same total number of activation and inhibition edges and as the biological network. The ER networks are fully randomized with respect to the number of activation and inhibition links per node – that is, they have no structural bias (*e.g.,* no hubs). By contrast, the SF networks share the same number of activation and inhibition links as the cell cycle network for each node, and therefore share the same degree distribution, defined as the rank ordering of the number of (in-directed and out-directed) edges per node. In what follows, we compare the informational properties of the fission yeast cell cycle biological network with samples drawn at random from both ER and SF network ensembles consisting of 1,000 sampled networks each, unless otherwise specified.

---

[2] In our study, Scale-Free (SF) network, unlike its common definition, does not mean that the sample networks in the ensemble exhibit power-law degree distributions. Instead, the name emphasizes that the sample networks have the same exact degree sequence as the fission yeast cell cycle and hence the networks share the same bias in terms of their global topology as the biological fission yeast cell-cycle network. For larger biological networks that are indeed truly scale-free, the analogous random graphs would therefore also be scale free.

We note that we purposefully do not use a fully randomized network ensemble having no structural features in common with the biological network in our comparison. This is because we are attempting to distinguish contributions to informational architecture peculiar to biological function from those that arise from more commonly studied topological features of the network, such as degree distribution. We note that many earlier studies focused solely on topological properties, such as scale-free degree distributions (where the rank ordering of the number of edges for each node follows a power-law), as distinctive aspects of evolved biological networks [17-19]. Our analysis of informational structure, however, uncovers a further (and potentially more significant) layer of distinctive features that go beyond topological considerations alone, which is best uncovered by considering random networks constrained to maintain topological features in common with the biological network.

# 3. Quantifying Informational Architecture

Information-theoretic approaches have provided numerous insights into the properties of distributed computation and information processing in complex systems [20]. Since we are interested in *level non-specific* patterns that might be intrinsic to biological organization, we investigate both local (node-to-node) and global (state-to-state) informational architecture. We note that in general, biological systems may often have more than two "levels", but focusing on two for the relatively simple case of the fission yeast cell cycle is a tractable starting point. To quantify local informational architecture we appeal to the information dynamics developed Lizier *et al.* [7-9], which utilizes Schreiber's transfer entropy [21] as a measure of information processing. For global architecture, we implement integrated information theory (IIT), developed by Tononi and collaborators, which quantifies the information generated by a network as a whole when it enters a particular state, as generated by its causal mechanisms [11]. In this section we describe each of these measures (more detailed descriptions may be found in [8] for information dynamics and [11] for integrated information theory). We note that while both formalisms have been widely applied to complex systems, they have thus far seen little application to direct comparison between biological and random networks, as we present here. We also note that this study is the first to our knowledge to combine these different formalisms to uncover informational patterns within the same biological network.

**3-1) Information Dynamics**

Information dynamics is a formalism for quantifying the local component operations of computation within dynamical systems by analysing time series data [7-9, 21]. We focus here on *transfer entropy* (TE) [21] and *active information* (AI), which are measures of information processing and storage respectively [8]. For the fission yeast cell cycle network, time series data is extracted by applying Eq. 1 for 20 time steps, for each of the 512 possible initial states, thus generating all possible trajectories for the network. Time series data were similarly generated for each instance of the random networks in our ensembles of 1,000 ER and SF networks. The trajectory length of t=20 time steps is chosen to be sufficiently long to capture transient dynamics for trajectories before converging on an attractor for the cell cycle and for the vast majority of random networks. Using this time series data, we then extracted the relative frequencies of temporal patterns and used these to define the probabilities *p* necessary to calculate TE and AI as discussed below (see *e.g.* [22] for an explicit example calculation of TE).

In an isolated mechanical system with no noise, past states are good predictors of future states. However, in a complex network, the past states of a given node will in general be inadequate to guide prediction of future states because of the strong nonlinear coupling of a node's dynamics to that of other nodes. Active information (AI) quantifies the degree to which uncertainty about the future of a node $X$ is reduced by knowledge of *only* the past states of that *same* node, found from examining time series data. Formally:

(2)

$$A_X(k) = \sum_{(x_n^{(k)}, x_{n+1}) \in \chi_1} p(x_n^{(k)}, x_{n+1}) \log_2 \frac{p(x_{n+1}, x_n^{(k)})}{p(x_n^{(k)}) p(x_{n+1})}$$

where $\chi_1$ indicates the set of all possible patterns of $(x_n^{(k)}, x_{n+1})$. Thus, AI is a measure of the mutual information between a given node's past, "stored" in its *k* previous states, and its immediate future (its next state).

An information-theoretic measure that takes into account the *flow* or transfer of information to a given node from other nodes in a network is transfer entropy (TE). It is the (directional) information transferred from a source node $Y$ to a target node $X$, defined as the reduction in uncertainty provided by $Y$ about the next state of $X$, over and above the reduction in uncertainty due to knowledge of the past states of $X$. Formally, TE from $Y$ to $X$ is the mutual information between the previous state of the source $y_n$ and the next state of the target $x_{n+1}$, conditioned on *k* previous states of target, $x^{(k)}$:

(3)

$$T_{Y \to X}(k) = \sum_{(x_n^{(k)}, x_{n+1}, y_n) \in \chi_0} p(x_n^{(k)}, x_{n+1}, y_n) \log_2 \frac{p(x_{n+1} | x_n^{(k)}, y_n)}{p(x_{n+1} | x_n^{(k)})}$$

where $\chi_0$ indicates the set of all possible patterns of sets of states $(x_n^{(k)}, x_{n+1, y_n})$. The directionality of TE arises due to the asymmetry in the computed time step for the state of the source and the destination. Due to this asymmetry, TE can be utilized to measure "information flows"[3], absent from non-directed measures such as mutual information.

Both TE and AI depend on the history length, *k*, which specifies the number of past state(s) of a given node one is interested in using to make predictions of the immediate future. Typically, one considers $k \to \infty$ (see *e.g.*, [9] for discussion). However, short history lengths can provide insights into how a biological network is processing information more locally in *space and time*. In particular, we note that it is unlikely that any physical system would contain infinite knowledge of past states and thus that the limit $k \to \infty$ may be unphysical. In truncating the history length $k$, we treat $k$ as a physical property of the network related to memory about past states as stored in its causal structure. TE enables us to quantify how much information is transferred between two distinct nodes at adjacent time steps, and thus provides insights into the *spatial* distribution of the information being processed. By contrast, we view AI as a measure of local information processed through *time*.

---

[3] Herein we use the term "information flow" to refer to the transfer of information between nodes. We use this interchangeably with the terminology "information transfer" and "information processing", this is not the same concept of information flow as proposed by Ay and Polani, which is a measure of causation [23].

We compared the distribution of AI and TE for the fission yeast cell cycle network with two types of randomized networks (ER and SF), averaging over ensemble of 500 and 1,000 networks, respectively. The results are shown in Fig. 3, and demonstrate that the biological network displays a significantly higher level of information transfer than either class of random networks on average.

Consider first the differences between the two comparison networks, ER and SF. Recall that ER networks do not share topological features with either the biological or SF networks except for the size of the networks, the number of self-loops and the total number of activation and inhibition links (Table 1). For the ER networks, connections between two nodes are made at random and as a result, the degree distribution is more homogeneous on average than that of the biological or SF networks (*e.g.*, most networks will not have hubs*)*. Our results indicate relatively low average information transfer (as measured by TE) between nodes in the ER networks (green, left panel Fig. 3) as compared to SF (blue, left panel Fig.3). The much higher TE between nodes in the biological and SF networks than the ER networks suggests that heterogeneity in the distribution of edges among nodes – which arises due to the presence of hubs in the SF and biological networks, but not the ER networks – plays a significant role in information transfer.

However, heterogeneity alone clearly does not account for the high level of information transfer observed between nodes in the fission yeast cell cycle network, which is distinguished from the ensemble of SF networks in Fig. 3 *despite sharing the same exact degree distribution.* We note that although scale free networks with power law degree distributions have been much studied in the context of metabolic networks [18], signaling and protein-protein networks within cells [24], functional networks within the brain [25] and even technological systems [19], there has been very little attention given to *how biological systems might stand out as different from generic scale free networks*. Here both the biological and SF networks share similarities in topological structure (inclusive of degree distribution). However, the cell cycle network exhibits statistically significant differences in the distribution of information processing among nodes. The excess TE observed in the biological network (red, left panel Fig. 3) deviates between $1\sigma$ to $5\sigma$ from that of the SF random networks (blue, left panel Fig. 3), with a trend of increasing divergence from the SF ensemble for lower ranked node pairs that still exhibit correlations (e.g., where TE > 0). Interestingly, the biologically distinctive regime is dominated by information transfer from other nodes in the network to the control kernel, and from the control kernel to other nodes, as reported in by us in [12]. This seems to suggest that the biologically significant regime is attributable to information transfer through the control kernel, which as noted in Section 2-2 also has been shown to regulate function. The biological network is also an outlier in the total quantity of information processed by the network, processing more information on average than either the ER or SF random null models: the network is in the 100[th] percentile for ER networks and 95[th] percentile for SF networks for total information processed [12].

For the present study we also computed the distribution of AI (information storage) for the biological network and random network ensembles (right panel in Fig. 3). Information storage can arise locally through a node's self-interaction (self-loop), or be distributed via causal interactions with neighboring nodes. For the biological network (red, right panel Fig. 3), the control kernel nodes (labeled in red on the x-axis) have the highest information storage. Control kernel nodes have no self-interaction, so their information storage must be distributed among direct causal neighbors. Nodes that are self-interacting (labeled in blue on the x-axis) tend to have relatively low AI by comparison to the control kernel nodes (in this network self-interaction models self-degradation of the represented protein). This suggests that the distribution of information storage in the fission yeast cell cycle arises primarily due to *distributed storage embedded within the network's causal structure*. This distributed information storage acts primarily to reduce uncertainty in the future state of the control kernel nodes, which have the highest information storage. We note that the patterns in information storage reported here are consistent with that reported in [26] since this network has inhibition links, which detract from information storage.

For the biological network, it is the control kernel nodes that store the most information about their own future state, as compared to other nodes in the network. The analogous nodes in the random networks also on average store the most information. Taken in isolation, our results for the distribution of AI therefore do *not* distinguish the biological network from random. However, we note that the random networks in our study are constructed to maintain features of the fission yeast cell cycle network's causal structure (see Table 1). It is therefore not so surprising that the ER and SF networks should share common properties in their information storage with the biological network. However, the interpretation of the distribution of AI among nodes is very different for the biological network than for the random networks. Why is this? The ensembles of random networks drawn from SF and ER networks will in general *not* share the same attractor landscape as the biological case (shown in Fig. 2). For the biological network, the control kernel nodes are associated with a specific attractor landscape associated with the function of the network. For the biological network, control kernel nodes contribute the most to information storage *and* are also associated with regulation of the dynamics of the biological network. For ER and SF ensembles, the analogous nodes likewise store a large amount of information (having inherited their *local* causal structure in the construction of our random ensembles), but these nodes do not necessarily play any role in regulating the *global* causal structure of the network. Thus although the AI patterns in Fig. 3 are not statistically distinct for the biological network as compared to the null models, only for the biological network is it the case that this pattern is associated with the function of the network, that is that *the nodes storing the most information via local causal structure also play a role in regulating the global causal structure*.

### 3-2) Effective and Integrated Information

Whereas information dynamics quantifies patterns in local information flows, integrated information theory (IIT) quantifies information arising due to the network's global properties defined by its state-to-state transitions (global causal structure) [11]. IIT was developed originally as a theory for consciousness; for technical details see e.g., [27], but is widely applicable to other complex systems. In this paper we use IIT to quantify *effective information* (EI) and *integrated information* ($\phi$) (both defined below) for the fission yeast cell cycle and also for the ensembles of random ER and SF networks. Unlike information dynamics, both EI and $\phi$ characterize the information generated by entire network *states*, rather than individual nodes. They do not require time series data to compute: one can calculate EI or $\phi$ for all causally realizable states of the network (states the network that are a possible output of its causal mechanisms*)* independent of calculating trajectories through state-space. These measures may in turn be mapped to the global causal structure of the network's dynamics through state space, *e.g.* for the fission yeast cell cycle, EI and $\phi$ for network states can be mapped to the network states in the flow diagram in Fig. 2.

Effective information (EI) quantifies the information generated by the causal mechanisms of a network (as defined by its edges, rules and thresholds – see e.g., Eq. (1)), when it enters a particular network state **G′**. More formally, the effective information for each realized network state **G′**, given by EI(**G′**), is calculated as the relative entropy (or Kullback-Leibler divergence) of the *a posteriori repertoire* with respect to the *a priori repertoire*:

(4) $\quad EI(\mathbf{G} \rightarrow \mathbf{G'}) = H\big(p^{max}(\mathbf{G})\big) - H(p(\mathbf{G} \rightarrow \mathbf{G'}))$

where *H* indicates entropy. The *a priori repertoire* is defined is as the maximum entropy distribution, $p^{max}(G)$, where all network states are treated as equally likely. The *a posteriori repertoire*, $p(G \to G')$, is defined as the repertoire of possible states that could have led to the state **G'** through the causal mechanisms of the system. In other words, EI(**G'**) measures how much the causal mechanisms of a network reduce the uncertainty about the possible states that might have preceded **G'**, *e.g.*, its possible causes. For the fission yeast cell cycle, the EI of a *state* is related to the number of in-directed edges (causes) in the attractor landscape flow diagram of Fig. 2.

*Integrated information* ($\phi$) captures how much the "the whole is more than the sum of its parts" and is quantified as the information generated by the causal mechanisms of a network when it enters a particular state **G'**, as compared to sum of information generated independently by its parts. More specifically, $\phi$ can be calculated as follows: 1) divide the network entering a state **G'** into distinctive parts and calculate EI for each part, 2) compute the difference between the sum of EIs from every part and EI of the whole network, 3) repeat the first two steps with all possible partitions. $\phi$ is then the minimum difference between EI from the whole network and the sum of EIs for its parts (we refer readers to [11] for more details on calculating $\phi$). If $\phi(G') > 0$, then the causal structure of network generates more information as a whole, than as a set of independent parts when it enters the network state *G'*. For $\phi(G') = 0$, there exist causal connections within the network that can be removed without leaking information.

The distribution of EI for all accessible states (states with at least one cause) in the fission yeast cell-cycle network is shown in Fig. 4 where it is compared to the averaged EI for the ER and SF null network ensembles. For the biological network, most states have *EI = 8*, corresponding to two possible causes for the network to enter that particular state (the *a priori* repertoire contains 512 states, whereas the *a posteriori* repertoire contains just two, so $EI(G \to G') = log_2(512) - log_2(2) = 8$ bits). Comparing the biological distribution to that of the random network ensembles *does not* reveal distinct differences (the biological network is within the standard deviation of the SF and ER ensembles), as it did for information dynamics. Thus, the fission yeast cell cycle's causal mechanisms do not statistically differ from SF or ER networks in their ability to generate effective information. Stated somewhat differently, the statistics over individual state-to-state mappings within the attractor landscape flow diagram for the fission yeast cell cycle (Fig. 2) and the ensembles of randomly sampled networks are indistinguishable.

Fig. 5 shows calculated values of $\phi$ for all network states that converge to the primary attractor of the fission yeast network (all network states within the red region in the left panel of Fig. 2). Larger points denote the biologically realized states, which correspond to those that a healthy functioning *S. Pombe* cell will cycle through during cellular division (*e.g.*, the states corresponding to the G1—S—G2—M phases). Initially, we expected that $\phi$ might show different patterns for states within the cell cycle phases, as compared to other possible network states (such that biologically functional states would be more integrated). However, the result demonstrates that there are *no* clear differences between $\phi$ for biologically functional states and other possible network states. We also compared the average value of integrated information, $\phi_{avg}$, taken over all realizable network states for the fission yeast cell cycle network, to that computed by the same analysis on the ensembles of ER and SF random networks. We found that there is no statistical difference between $\phi_{avg}$ for the biological and random networks: as shown in Table 2, all networks in our study show statistically similar averaged integrated information.

At first, we were surprised that neither EI nor $\phi$ (or $\phi_{avg}$) successfully distinguished the biological fission yeast cell cycle network from the ensembles of ER or SF networks. It is widely regarded that a hallmark of biological organization is that "more is different" [28] and that it is the emergent features of biosystems that set them apart from other classes of physical systems [29]. Thus, we expected that global properties would be more distinctive to the biological networks than local ones. However, for the analyses presented here, this is not the case: the local informational architecture, as quantified by TE, of biological networks is statistically distinct from ensembles of random networks: yet, their global structure, as quantified by EI and $\phi$, is not. There are several possible explanations for this result. The first is that we are not looking at the "right" biological network to observe the emergent features of biological systems. While this may be the case, this type of argument is not relevant for the objectives of the current study: if biological systems represent physics best captured by informational structure, than one should not be able to cherry-pick which biological systems have this property – it should be a universal feature of biological organization. Hence our loose analogy to the "hydrogen atom" – given the universality of the underlying atomic physics we would not expect helium to have radically different physical structure than hydrogen does. Thus, we expect that if this type of approach is to have merit, the cell cycle network is as good a candidate case study as any other biological system. We therefore consider this network as representative, given our interest in constraining what *universal physics* could underlie biological organization. One might further worry that we have excised this particular network from its environment (*e.g.*, a functioning cell, often suggested as the indivisible unit, or "hydrogen atom of life"). This kind of excision might be expected to diminish emergent informational properties – it is then perhaps more surprising that the *local* signature in TE remains so prominent even though EI and $\phi$ are not statistically distinct.

Another possible explanation for our results is that we have defined our random networks in a way that makes them too similar to the biological case, thus masking some of the differences between functional biological networks and our random network controls. It is indeed likely that biologically-inspired "random" networks will mimic some features of biology that would be absent in a broader class of random networks (*e.g.* such as the specific pattern in the distribution of information storage discussed in the previous section). However, if our random graph construction is too similar to the biological networks to pick up important distinctive features of biological organization, than it does not explain the observed unique patterns in TE nor that AI is largest for control kernel nodes which play a prominent role in the regulation of function for the biological network. We therefore accept that the lack of distinct emergent, global patterns in information generated due to causal architecture as a real feature of biological organization and not an artifact of our construction. This observation may offer clues to what may in fact be the most distinct feature of biological systems.

# 4. Characterizing Informational Architecture

The forgoing analyses indicate that what *distinguishes biology as a physical system cannot be causal structure (topology) alone, but instead biology can be distinguished by its informational architecture*, which arises as an emergent property of the combination of *topology and dynamics*. This is supported by the distinct scaling in information processing (TE) observed for the fission yeast cell cycle network as compared to ER and SF ensembles reported above. In our example, what distinguishes biology from other complex physical systems cannot be global topological features alone, since the SF networks differ in their patterns of information processing from the fission yeast cell cycle, despite sharing common topological features, such as degree distribution. Similarly, it cannot be dynamics alone due to the lack of a distinct signature in EI or $\phi$ for the biological network, as both are manifestations the global dynamics on the attractor landscape.

An important question opened by our analysis is why the biological network exhibits distinct features for TE and shows patterns in AI associated with functional regulation, when global measures yield no distinction between the biological network and random networks. We suggest that the separate analysis of two distinct levels of informational patterns (*e.g.* node-node *or* state-state) as presented above misses what arguably may be one of the most important features of biological organization – that is, that *distinct levels interact*. This view is supported by the fact that the control kernel plays a prominent role in the distinctive local informational patterns of the fission yeast cell cycle network, but was in fact first identified for its role regulating dynamics on the global attractor landscape. The control kernel may therefore be interpreted as *mediating the connection between the local and global causal structure of the network*. The results of the previous section indicate that these same nodes act as a hub for the transfer of information within the network and for information storage. The interaction between distinct levels of organization is typically described as 'top-down' causation and has previously been proposed as a hallmark feature of life [30-32]. We hypothesize that the lower level patterns observed with TE and AI arise because of the particular manner in which the biological fission yeast cell cycle network is integrated, regulating dynamics on its global attractor landscape through the small subset of control kernel nodes via `top-down' control. Instead of studying individual levels of architecture as separate entities, to fully understand the informational architecture of biological networks we must therefore additionally study informational patterns mediating the *interaction*s between different levels of informational structure, as they are distributed in space and time. To test this hypothesis we analyse the spatiotemporal and inter-level architecture of the fission yeast cell cycle.

### 4-2) Spatiotemporal Architecture

We may draw a loose analogy between information and energy. In a dynamical system, energy is important to defining two characteristic time scales: a dynamical process time (*e.g.,* the period of a pendulum) and the dissipation time (*e.g.,* time to decay to an equilibrium end state or attractor.) TE and AI are calculated utilizing time series data of a network's dynamics and, as with energy, there are also two distinct time scales involved: the history length *k* and the convergence time to the attractor state(s), which are characteristic of the dynamical process time and dissipation time, respectively. For example, the dissipation of TE may correlate with biologically relevant time scales for the processing of information, which can be critical for interaction with the environment or other biological systems. In the case study presented here, the dynamics of TE and AI can provide insights into the timescales associated with information processing and memory within the fission yeast cell cycle network.

The TE scaling relation for the fission yeast cell cycle network is shown in Fig. 6 for history lengths *k = 1, 2 … 10* (note: history length *k = 2* is compared to random networks in Fig. 2). The overall magnitude and nonlinearity of the scaling pattern decreases as knowledge about the past increases with increasing *k*. The patterns in Fig. 6 show general trends that indicate that the network processes less information in the spatial dimension when knowledge about past states increases. In contrast, the temporal processing of information, as captured by information storage (AI) increases for increased *k* (not shown).

To make explicit this trade-off between information processing and information storage we define a new information theoretic measure. Consider *in-coming TE*, and *out-going TE* for each node as the total sum of TE from the rest of network to the node and the total sum of TE from the node to the rest of network, respectively. We then define the *Preservation Entropy* (PE), as follows:

(5) $$P_X(k) = A_X(k) - \tfrac{1}{2}\bigl(T_X^I(k) + T_X^O(k)\bigr)$$

where $A_X(k)$, $T_X^I(k)$ and $T_X^O(k)$ denote AI, in-coming TE, and out-going TE, respectively. PE quantifies the difference between the information stored in a node and the information it processes. For PE(X) > 0, a node **X**'s temporal history (information storage) dominates its information dynamics, whereas for PE(X) < 0, the information dynamics of node **X** are dominated by spatial interactions with rest of the network (information processing). Preservation entropy is so named because nodes with PE > 0 act to *preserve* the dynamics of their own history.

Fig. 7 shows PE for every node in the fission yeast network, for history length *k = 1, 2 … 10*. As the history length increases the overall PE also increases, with all nodes acquiring positive values of PE for large *k*. For all *k*, the control kernel nodes have the highest PE, with the exception of the start node SK (which only receives external input). When the dynamics of the cell cycle network is initiated, knowledge about future states can only be stored in the spatial dimension so PE < 0 for all nodes. The transition to temporal storage first occurs for the four control kernel nodes, which for *k = 4* have PE > 0, while others nodes have negative PE (self-loops nodes) or PE close to 0. All nodes make the transition to PE > 0 at history length *k =*

5. This suggests that the informational architecture of the cell cycle is *space (processing)* dominated in its early evolution and later transitions to being *time (storage)* dominated with a characteristic time scale.

### 4-3) Inter-level Architecture

We pointed out in Section 3 the fact that the local level informational architecture picks out a clear difference between biological and random networks, whereas the global measures do not. We conjecture that this is because in biological systems important information flows occur *between* levels, *i.e.* from local to global and vice-versa, rather than being solely characteristic of local or global organization alone. In other words, network integration may not distinguish biological from random, but the particular manner in which a network is integrated and how this filters to regulate lower level interactions may be distinctive of life. In the case study presented here, this means there should be distinctive patterns in information flows arising from node-state, or state-node interactions.

To investigate node-state and state-node interactions, we treat $\phi$ itself as a dynamic time series and ask whether the dynamical behavior of individual nodes is a good predictor of $\phi$ (*i.e.*, of network integration), and conversely, if network integration enables better prediction about the states of individual nodes. To accomplish this we define a new Boolean "node" in the network, named the *Phi-node*, which encodes whether the state is integrated or not, by setting its value to 1 or 0, respectively (in a similar fashion to the mean-field variable in [30]). We then measure TE between the state of the Phi-node and individual nodes for all possible trajectories of the cell cycle, in the same manner as was done for calculating TE between local nodes. Although transfer entropy was not designed for analyses of information flows between 'levels', there is actually nothing about the structure of any of the information measures utilized herein that suggests they must be level specific (see [33] for an example of the application of EI at different scales of organization). Indeed, higher transfer entropy from global to local scales than from local to global scales has previously been put forward as a possible signature of collective behavior [31, 34-36].

The results of our analysis are shown in Fig. 8. The total information processed (total TE) from the global to local scale is 1.6 times larger than total sum of TE from the local to global scale. That is, the cell cycle network tends to transfer more information from the global to local scales (top-down) than from the local to global (bottom-up), indicative of collective behavior arising due to network integration. Perhaps more interesting is the irregularity in the distribution of TE among nodes for information transfer from global to local scales (shown in purple in Fig. 8), as compared to a more uniform pattern in information transfer from local to global scales (shown in orange in Fig. 8). This suggests that only a small fraction of nodes act as optimized channels for filtering globally integrated information to the local level. This observation is consistent with what one might expect if global organization is to drive local dynamics (*e.g.*, as is the case of top-down causation), as this must ultimately operate through the causal mechanisms at the lower level of organization (such that it is consistent with known physics). Our analysis suggests a promising line of future inquiry characterizing how the integration of biological networks may be structured to channel global state information through a few nodes to regulate function (*e.g.*, such as the control kernel). Future work will include comparison of the biological distribution for TE between levels to that of random network models to gain further insights into if, and if so how, this feature may be distinctive to biology. We note, however, that this kind of analysis requires one to regard the level of integration of a network as a "physical" node. But perhaps this is not too radical a step: effective dynamical theories often treat mean-field quantities as physical variables.

## 5. Discussion

Several measures of both information storage and information flow have been studied in recent years, and we have applied these measures to a handful of tractable biological systems. Here we have reported our results on the regulatory network that controls the cell cycle of the fission yeast *S. Pombe*, treated as a simple Boolean dynamical system. We confirmed that there are indeed informational signatures that pick out the biological network when contrasted with suitably defined random comparison networks. Intriguingly, the relevant biosignature measures are those that quantify local information transfer and storage (TE and AI respectively), whereas global information measures such as integrated information $\phi$ do not. The distinguishing feature of these measures is that information dynamics is *local and correlative* (using conditional probabilities), whereas integrated information is *global and causative* (using *interventional* conditional probabilities [37]). The signature of biological structure uncovered by our analysis therefore lies not with the underlying causal structure (the network topology) but with the informational architecture via specific patterns in the distribution of correlations unique to the biological network.

Although the biological fission yeast cell cycle network and ensemble of random networks in our study share commonalities in causal structure, the pattern of information flows for the biological network is quite distinct from either the SF or ER random networks. We attribute this difference to the presence of the control kernel nodes. Both the biological network and the ensemble of SF random networks statistically differ in the distribution of the transfer of information between pairs of nodes from the more generic ER random networks. Surprisingly, the newly uncovered scaling relation in information transfer is statistically distinct for the biological network, even among the class of SF networks sharing a common degree distribution. The biologically most distinct regime of the scaling relation is associated with information transfer to and from the control kernel nodes and the rest of the network as reported by us in [12]. Our results presented here indicate the cell cycle informational architecture is structured to localize information storage within those same control kernel nodes. These results indicate that the control kernel – which plays a prominent role in the regulation of cell cycle function (by regulating the attractor landscape) – is a key component in the distinctive informational architecture of the fission yeast cell cycle network, playing both a prominent role in information storage and in the flow of information within in the network.

While it is conceivable that these patterns are a passive, secondary, attribute of biological organization arising via selection on other features (such as robustness, replicative fidelity *etc*), we think the patterns are most likely to arise because they are intrinsic to biological function – that is, they direct the causal mechanisms of the system in some way, and thereby constitute a directly selectable trait [38; 39]. If information does in fact play a causal role in the dynamics of biological systems, than a fundamental understanding of life as a physical process has the potential to open up completely unexplored sectors of physics, as we know of no other class of physical systems where information is necessary to specify its state. Taking a more forward-thinking and necessarily speculative look at what our results suggest of the physics underlying life, we regard the most distinctive feature to be in how informational and causal structure intersect (consistent with other suggestions that life is distinguished by the "active" use of information, see *e.g.*, [22, 31, 41-43]). Evidence for this view comes from the fact that the integration of the network is a better predictor of the states of individual nodes, than vice versa (see Fig. 8), an asymmetry perhaps related to the functionality of the network. If the patterns of information processing observed in the biological network are indeed a joint product of information and causal structure, as our results suggest, they may be regarded as an *emergent property of topology and dynamics*. The informational signatures of biological networks uncovered by our analysis appear strongly dependent on the controllability of the network – that is, that a few nodes regulate the function of the cell cycle network. Thus, *in addition to scale free network topology, a necessary feature required to distinguish biological networks, based on their informational architecture, is the presence of a subset of nodes that can "control" the dynamics of the network on its attractor landscape*. In this respect, biological information organization differs from other classes of collective behaviour commonly described in physics. In particular, the distribution of correlations indicates that we are not dealing with a critical phenomenon in the usual sense, where correlations at all length scales exist in a given physical system at the critical point, without substructure [40]. Instead, we find "sub-critical" collective behavior, where a few network nodes centralize correlations and regulate collective behavior through the global organization of information flows.

The control kernel nodes were first discovered by pinning their values to that of the primary attractor – that is, by causal intervention (see *e.g.,* [37]). However, causal intervention by an *external* agent does not occur "in the wild". We therefore posit that the network is organized such that information flowing through the control kernel performs an analogous function to an external causal intervention. Indeed, this is corroborated by our results demonstrating that the network transitions from being information "processing" to "storage" dominated: the control kernel nodes play the dominant role in information storage for all history lengths $k$ and are the first nodes to transition to storage-dominated dynamics. We additionally note that the control kernel state takes on a distinct value in each of the network's attractor states, and thus is related to the *distinguishability* of these states, as recognized in Kim *et al* [13]. Based on our results presented here, a consistent interpretation of this feature is that the control kernel states provide a "coarse-graining" of the network state-space relevant to the network's function that is *intrinsic* to the network (not externally imposed by an observer). Storing information in control kernel nodes therefore provides a physical mechanism for the network to internally manage information about its own global state space. This interpretation is also consistent with Kim *et al*'s observations that the size of the control kernel scales both with the number and size of attractors. In short, one interpretation of our results is that the network is organized such that information processing that occurs in early times "intervenes" on the state of the control kernel nodes, which in turn transition to storage-dominated dynamics that regulate the network's dynamics along the biologically functional trajectory. If true, *biology may represent a new frontier in physics where information (via distributed correlation in space and time) is organized to direct causal flows*. The observed scaling of transfer entropy may be a hallmark of this very organization and therefore a universal signature of regulation of function, and thus life.

## Data Accessibility
There is no supplementary material for this manuscript.

## Competing Interests
The authors have no competing interests.

## Author Contributions
HK carried developed the computational algorithms for constructing random networks and for calculating information measures, performed data analysis, and helped draft the manuscript. SW conceived of the study, designed the study, coordinated the study, contributed to data interpretation and drafted the manuscript. PD contributed to data interpretation and critically revised the manuscript. All authors gave final approval for publication.

## Acknowledgements
This project was made possible through support of a grant from Templeton World Charity Foundation. The opinions expressed in this publication are those of the author(s) and do not necessarily reflect the views of Templeton World Charity Foundation. The authors wish to thank Larissa Albantakis, Joseph Lizier, Mikhail Prokopenko, Paul Griffiths, Karola Stotz and the Emergence@ASU group for insightful conversations regarding this work.

## Funding Statement

The authors were supported by a grant from the Templeton World Charity Foundation.

**Tables**

|  | **Erdös-Rényi (ER) networks** | **Scale-Free (SF) networks** |
|---|---|---|
| **Size of network (Total number of nodes, inhibition and activation links)** | Same as the cell-cycle network | Same as the cell-cycle network |
| **Nodes with a self-loop** | Same as the cell-cycle network | Same as the cell-cycle network |
| **The number of activation and inhibition links for each node** | **NOT** the same as the cell-cycle network (➔ no structural bias) | Same as the cell-cycle network (➔ Same degree distribution) |

Table 1. Constraints for constructing random network graphs that retain features of the causal structure of a reference biological network, which define the two null model network classes used in this study: Erdos-Renyi (ER) networks and Scale-Free (SF) networks.

| Network Type | $\phi_{avg}$ | $\Delta \phi$ |
|---|---|---|
| Fission Yeast Cell-Cycle Network | 0.151 | 0 |
| ER Random Networks | 0.099 | 0.008 |
| SF Random networks | 0.170 | 0.004 |

Table 1. Comparison of the state-averaged integrated information for the fission yeast cell-cycle network

and two types of null model networks reveals no statistically significant differences for the biological network.

**Figures**

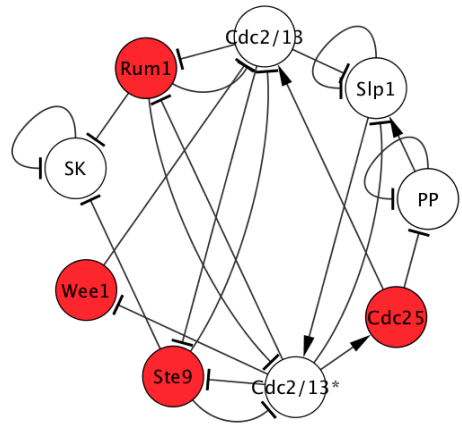

Figure 1. Boolean network model representing interactions between proteins regulating cell cycle function in the fission yeast *S. Pombe*. Nodes and edges represent regulatory proteins and their causal interactions, respectively. Arrows and bars represent activation and inhibition links, respectively. Control kernel nodes regulating the function of the cell cycle network are highlighted in red. Figure adopted from [13].

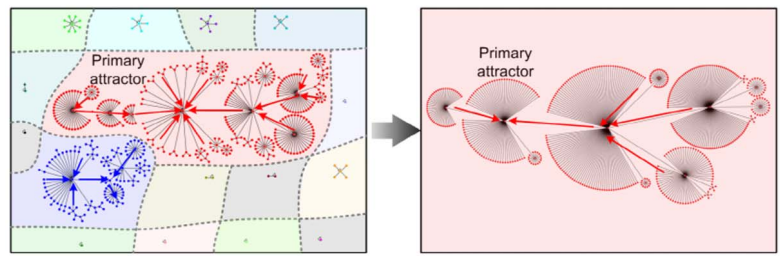

Figure 1. Attractor landscape of the fission yeast cell cycle network, showing all possible dynamic trajectories of network states. Left panel: Attractor landscape for the fission yeast cell cycle regulatory network. Regions are coloured by the attractor to which states within that region converge. The attractor landscape is dominated by the primary attractor associated with biological function shown in red, which includes ~70% of network states. Right panel: The attractor landscape is changed after global regulation caused by pinning the values of control kernel notes to their state in the primary attractor: all network states converge to the primary attractor under regulation by the control kernel. Figure adopted from [13].

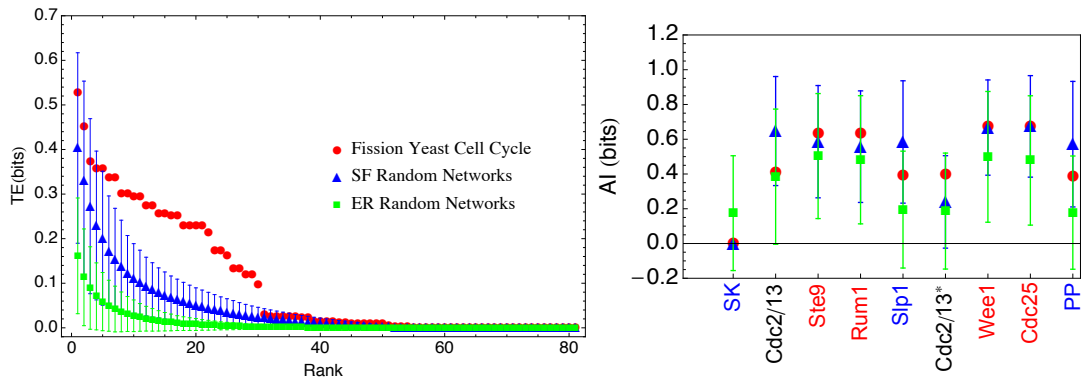

Figure 3. Scaling distribution of information processing among node pairs (measured with TE) and the information storage for individual nodes (measured with AI). Shown are results for the fission yeast cell cycle regulatory network (red) and ensembles of Erdos-Renyi (ER) random networks (green) and Scale Free (SF) random networks (blue). Left panel: Scaling of information processing for history length k= 2, shows that the biological network processes more information than either ER or SF networks on average. The y-axis and x-axis are the TE between a pair of nodes and relative rank, respectively. Ensemble statistics are taken over a sample of 1,000 networks. Figure adopted from Kim et al. [12]. Right Panel: AI for all nodes for history length k = 5. Ensemble statistics are taken over a sample of 500 networks. Nodes names correspond to regulatory proteins for the biological cell cycle network, with control kernel nodes highlighted in red and nodes with a self-loop in blue. For the random networks, the labels are retained to indicate nodes in the network relative to the cell cycle network, but do not correspond to real proteins as is the case for the cell cycle network. Although not statistically distinct since control kernel nodes and their analogs in the random networks store the most information, only in the biological network is this information storage associated with regulation of they dynamics of the attractor landscape.

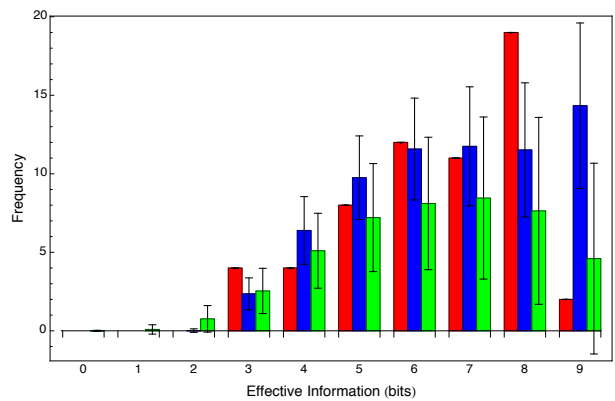

Figure 4. Distribution of effective information (EI) for the fission yeast cell-cycle regulatory network and ensembles of Erdos-Renyi (ER) random networks (green) and Scale Free (SF) random networks (blue). Values are calculated for every state each network can enter via its causal mechanisms. The data shows that the distribution of EI for biological network is not statistically distinct from random.

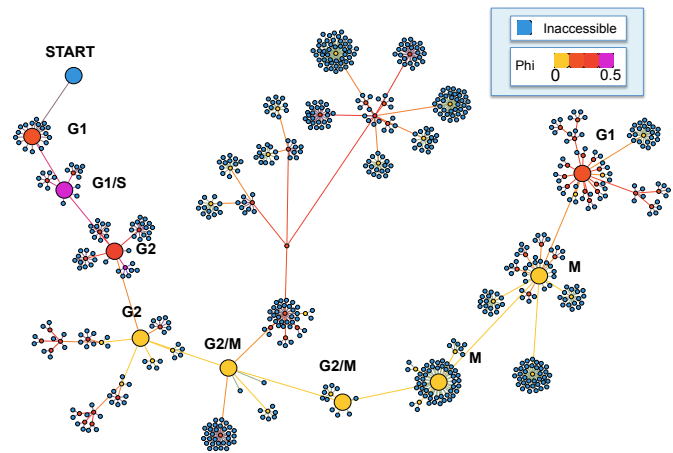

Figure 5. Diagram illustrating the integrated information of each network states in the primary basin of attraction for the fission yeast cell-cycle regulatory network. Colours indicate the value of integrated information for each state. Large points represent states in the functioning (healthy) cell cycle sequence, which do no show significant differences in terms of their integration from other network states.

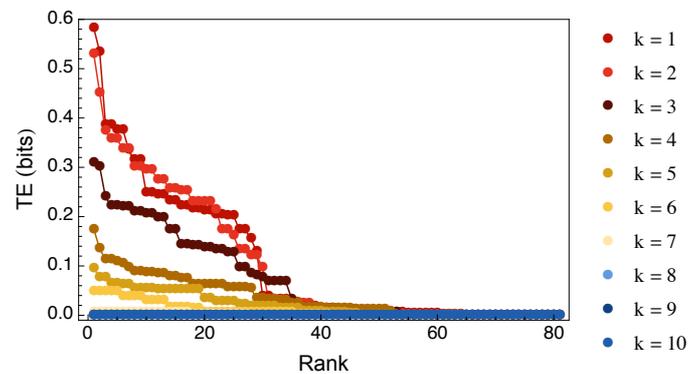

Figure 6. Scaling of information transfer (as measured by TE) for every pair of nodes in the fission yeast cell-cycle network, shown for history lengths $k = 1, 2, \ldots 10$ (results for $k=2$ are shown in Fig. 4 contrasted with ensembles of Erdos-Renyi and Scale Free null model networks). Information processing is high for short history lengths, but rapidly "dissipates".

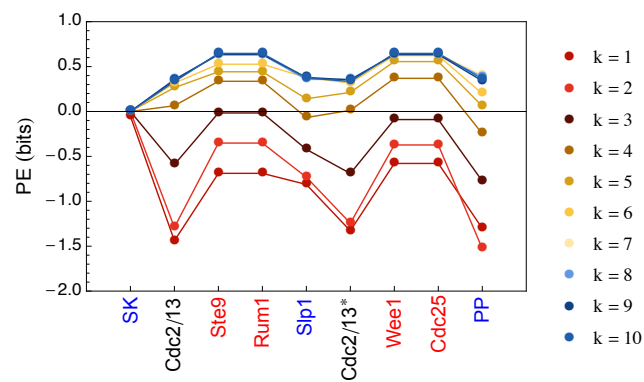

Figure 7. Preservation entropy (PE), quantifying the difference between information storage and transfer for a node, for every node in the fission yeast network for history lengths $k = 1, 2, \ldots 10$. Red and blue coloured nodes represent control kernel nodes and nodes with a self-loop, respectively. Control kernel

nodes have the highest PE for any history length, and transition from being processing dominated (PE < 0) to storage dominated (PE>0) first.

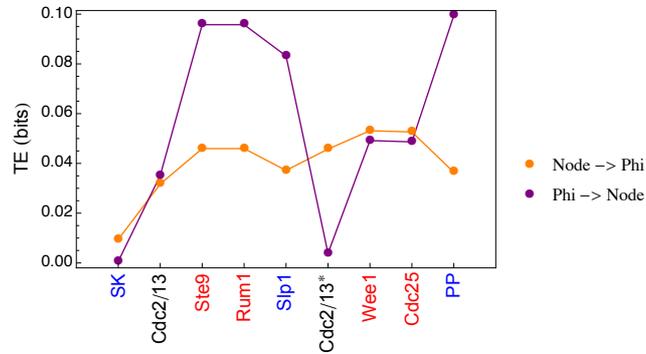

Figure 8. Information transfer between network integration, as quantifed by $\phi$ for individual states (the "Phi-node"), and individual nodes in the fission yeast cell cycle regulatory network. The orange line shows the TE from node → network state (individual nodes to the Phi-node) and the purple line shows TE from network state → node (Phi node to individual nodes). The network transfers 1.6 times more information from global to local scales than vice versa, indicative of collective behaviour, and this information transfer is asymmetrically distributed among nodes.